\documentclass[sigconf,natbib=true]{acmart}
\AtBeginDocument{%
  }

\setcopyright{acmlicensed}
\copyrightyear{2026}
\acmYear{2026}
\acmDOI{XXXXXXX.XXXXXXX}

\acmConference[Conference acronym 'XX]{Make sure to enter the correct
  conference title from your rights confirmation email}{June 03--05,
  2018}{Woodstock, NY}

\begin{document}

\title{Personalized Recommender Systems for Gym Workouts: A Reinforcement Learning Approach}

\author{Roan Rosema}
\affiliation{%
  \institution{Delft University of Technology}
  \city{Delft}
  \country{The Netherlands}}
\email{r.rosema@student.tudelft.nl}

\author{Helma Torkamaan}
\affiliation{%
  \institution{Delft University of Technology}
  \city{Delft}
  \country{The Netherlands}}
\email{h.torkamaan@tudelft.nl}

\author{Masoud Mansoury}
\affiliation{%
  \institution{Delft University of Technology}
  \city{Delft}
  \country{The Netherlands}}
\email{m.mansoury@tudelft.nl}

\begin{abstract}
  Workout recommender systems aim to help gym users complete effective and engaging training sessions. However, recommending exercises alone is insufficient, as a practical system must also determine appropriate sets, repetitions, and training loads, while adapting to user behavior such as skipping exercises. Existing approaches typically consider only a subset of these factors, limiting their applicability in real-world settings. In this paper, we extend workout recommendation from exercise selection to full workout prescription. We propose a reinforcement learning (RL)-based framework with four environments: exercise-only and full-prescription settings, each with and without skip-based interaction. The full-prescription environments recommend exercises, sets, repetitions, and load, while the skip-enabled environments use user skipping behavior for online personalization. Experiments with synthetic users show that modeling the full prescription task leads to higher rewards and greater user engagement than exercise-only recommendation, highlighting the importance of realistic workout planning in personalized gym recommender systems.
\end{abstract}




\maketitle

\section{Introduction}

Regular physical activity is associated with better health outcomes across the lifespan, and current public-health guidelines recommend both aerobic activity and muscle-strengthening activity as part of a healthy lifestyle \cite{who2020pa_guidelines}. Fitness recommendation is therefore an impactful health-recommender domain: recommendations can affect adherence, safety, and long-term wellbeing, not only short-term item satisfaction \cite{Smyth_2019, evaluating_efficacy_10.1145/3308557.3308716}.

In gym training specifically, deciding what a person should do is far from straightforward. A useful recommendation is not only about which exercise to perform, but also about how it should be prescribed, including sets, repetitions, load, rest, and progression over time. These choices depend on the user's goal, training status, and current physical capacity~\cite{acsm2009progression}. We refer to this richer formulation as the \textit{full-prescription} setting, which captures all of the critical factors required for practical deployment.

Despite meaningful progress in workout recommendation, most existing work addresses only a subset of these factors. For instance, \citet{keeping_ijcai2023p692} focuses on aspects such as goal alignment, intra- and inter-session diversity, and fatigue, while \citet{recsys_with_ai_for_fas_8441895} considers parts of prescription, such as recommending exercises, sets, repetitions, and rest times. These approaches, however, fall short of the complexity found in real gym environments, where effective prescription demands the joint consideration of multiple interdependent variables.

To address this gap, we propose a reinforcement learning (RL) framework for gym workout recommendation that operates in the full-prescription setting. We formulate session construction as a sequential decision-making problem, in which an agent builds a workout step by step based on the user's goals, training status, current capacity, and feedback from prior sessions. Rather than selecting exercises alone, the agent prescribes sets, repetitions, and load alongside each recommendation. Crucially, the agent observes user feedback, including skipped exercises, and adapts its subsequent recommendations accordingly, making the system progressively more attuned to individual needs.

This paper makes three main contributions. First, we formulate single-session gym workout recommendation as a sequential prescription problem, introducing both exercise-only and full-prescription environments. Second, we introduce skip-enabled environments in which skip-only feedback drives online personalization during session construction. Third, we provide a controlled comparison between RL and non-RL baselines across static, dynamic, and stress-test user conditions, showing that RL offers the greatest advantage when the system must jointly determine exercise selection and user-specific dosage. Experiments with synthetic users demonstrate the superiority of our framework over baselines on both engagement and prescription quality.

\section{Related work}

Personalized physical activity recommender systems sit between standard recommender systems, sensing, coaching, and behavior change. The task is not to pick an item, but to recommend an activity or session that fits a user's context, goals, ability, and constraints \cite{running_recommendations_10.1145/3209219.3209269,Smyth_2019,recfit_10.1145/2676431.2676439}. In this domain, offline accuracy alone is often insufficient, because recommendations affect bodies and behavior, and systems must consider adherence, safety, motivation, and long-term engagement \cite{evaluating_efficacy_10.1145/3308557.3308716,investigating_Coppens2025}.

Prior work has studied several parts of this broader problem, including endurance-sport recommendation \cite{pace_my_race_10.1145/3298689.3346991,fit_to_run_10.1145/3383313.3412228}, context-aware and privacy-aware fitness personalization \cite{privacy-perserving_10.1145/3572899,a_rec_approach_Sanchez2020}, longitudinal physical activity recommendation \cite{balancing_habit_repetition_10.1145/3640457.3691715,repeating_my_workouts_10.1145/3631700.3664867}, and exercise-form feedback using sensors or computer vision \cite{fitcoach_8057208,using_learnable_physics_10.1145/3604915.3608816,lift_it_up_right_10.1145/3705328.3759314}. These works show that health and fitness recommendation is more structured than standard item ranking.

Reinforcement learning is relevant when recommendations unfold sequentially and the value of an action depends on later outcomes \cite{afsar2022rlsurvey,sutton2018reinforcement}. The closest prior work is the home-fitness RL framework of Tragos et al. \cite{keeping_ijcai2023p692}, where each episode is a workout session composed of bodyweight exercises. We build on this framework but change the setting in three ways: from home workouts to gym workouts, from exercise-only recommendation to full prescription, and from limited interaction modeling to explicit skip-only feedback with online personalization.

\section{Our proposed framework}
We model gym workout recommendation as an episodic RL problem. The agent acts as a workout planner that constructs a session step by step. Each episode represents a single workout session with fixed length $T=8$. The session-level focus is important: the agent is not optimizing a full training program across weeks, but a single workout whose quality depends on exercise order, accumulated workload, prescription quality, and possible user skipping.

The task can be formalized as an episodic Markov Decision Process (MDP) ~\cite{WHITE19891, sutton2018reinforcement}. At each step $t \in \{1,\dots,T\}$, the agent receives an observation vector $o_t$, selects an action $a_t \sim \pi_\theta(\cdot \mid o_t)$, and the environment transitions to a new state. Some user features are hidden from the agent, such as a true capacity value that affects skipping. For that reason, the problem is more naturally viewed as a partially observable MDP. In practice, we treat it as an MDP over the observation space by including user features, session history, and interaction signals in $o_t$.

We consider two action formulations. In the exercise-only setting, the agent selects a single exercise index from a catalog of size $N$. In the full-prescription setting, the agent selects both the exercise and prescription parameters:
\begin{equation}
    a_t = \big(e_t,b^{\text{set}}_t,b^{\text{rep}}_t,b^{\text{load}}_t\big),
\end{equation}
where $e_t \in \{0,\dots,N-1\}$ and $b^{\text{set}}_t$, $b^{\text{rep}}_t$, and $b^{\text{load}}_t$ are bin indices for sets, repetitions, and load, respectively. The full-prescription setting changes the nature of the problem. The agent must no longer only decide which exercise comes next, but also how strongly that exercise should be dosed for the current user. A good exercise paired with an unrealistic load or repetition range is a poor gym recommendation. In the final implementation, the exercise-only environments have $217$ possible actions per step. The full-prescription environments use $5$ set bins, $19$ repetition bins, and $21$ load bins, giving $217 \times 5 \times 19 \times 21 = 432{,}915$ possible action combinations per step. Based on the action-space choices and interaction settings, we define four environments, summarized in Table~\ref{tab:environment_overview}.

\begin{table}[t]
\centering
\small
\caption{Overview of the four environments. EONS = ExerciseOnlyNoSkip, EOS = ExerciseOnlySkip, FPNS = FullPrescriptionNoSkip, and FPS = FullPrescriptionSkip. Skip-enabled environments use completion scaling and online personalization.}
\label{tab:environment_overview}
\vspace{-10pt}
\begin{tabular}{lccc}
\toprule
Environment & Action space & Skips & Reward basis \\
\midrule
EONS & Discrete$(N)$ & No & Suggested session \\
EOS & Discrete$(N)$ & Yes & Completed session \\
FPNS & MultiDiscrete$(N,5,19,21)$ & No & Suggested session \\
FPS & MultiDiscrete$(N,5,19,21)$ & Yes & Completed session \\
\bottomrule
\end{tabular}
\end{table}

\begin{figure*}[t]
    \centering
    \includegraphics[width=\textwidth]{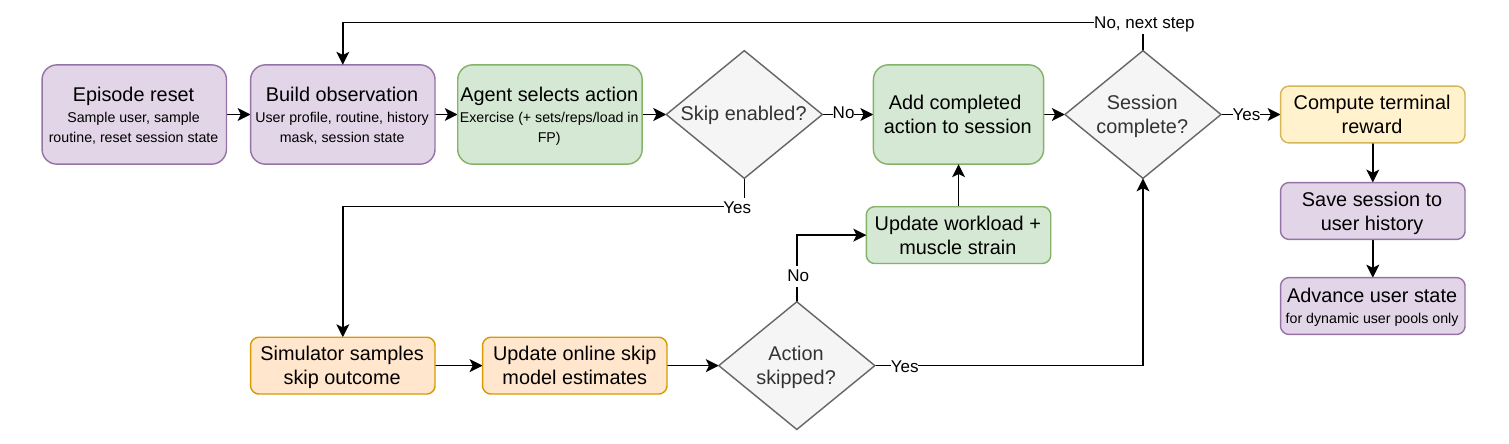}
    \vspace{-25pt}
    \caption{Simulation process for one episode. The agent constructs a session step by step. In skip-enabled environments, the simulator samples skip outcomes and the online skip model is updated from skip-only feedback. After the session is complete, a terminal reward is computed, the session is saved to user history, and dynamic user pools may advance the user state.}
    \label{fig:simulation-process}
\end{figure*}

Figure~\ref{fig:simulation-process} shows how one simulated episode unfolds. At each step, the agent proposes either an exercise or a full prescription. No-skip environments treat every action as completed, while skip-enabled environments sample completion and update online user estimates from skip-only feedback. After the session ends, a terminal reward is computed, the session is saved to user history, and dynamic user pools may advance the user state. The simulator and skip model are described in more detail in Section~\ref{sec:datasets_setup}.

All environments use a terminal, session-level reward. The step reward $r_t$ is zero for all intermediate steps, and a non-zero reward is returned only after the full session has been constructed:
\begin{equation}
    r_t=0 \quad \text{for } t<T, \qquad r_T=R(\tau).
\end{equation}
The reward combines sequence-level qualities such as uniqueness, intra-session diversity, inter-session diversity, routine alignment, and fatigue management. Full-prescription environments additionally include prescription quality and persistence. Skip-enabled environments compute the reward on the completed session and scale it by the completion ratio:
\begin{equation}
    R_{\text{final}} = c \cdot R_{\text{total}}, \qquad c=\frac{|S|}{T}.
\end{equation}
\noindent where $R_{\text{final}}$ is the terminal reward returned by the environment, $R_{\text{total}}$ is the weighted reward before completion scaling, $|S|$ is the number of completed exercises within a session (e.g., $|S|=6$ if 6 out of 8 exercises are completed), and $c$ is the completed fraction of the planned session. For a set of reward components $\{R_k\}_{k\in\mathcal{K}}$ where $\mathcal{K}$ is the set of all reward components for an environment, each component has a threshold $\tau_k$ and weight $w_k$. Components below their threshold are replaced by a fixed penalty before weighting:
\begin{equation}
\tilde{R}_k =
\begin{cases}
-0.5 & \text{if } R_k < \tau_k,\\
R_k & \text{otherwise},
\end{cases}
\qquad
R_{\text{total}}=\mathrm{clip}_{[-1,1]}\left(\sum_{k\in\mathcal{K}} w_k\tilde{R}_k\right).
\end{equation}
\noindent where $\mathrm{clip}_{[-1,1]}(x)$ clips $x$ to the interval $[-1,1]$. The detailed component definitions, thresholds, weights, and reward-design choices are provided in the supplementary material\footnote{Supplementary materials are available at \url{https://anonymous.4open.science/r/RL-RecSys-for-Gym-Exercises-FINAL-anon/}}, because the reward structure strongly influences both the learned policy and the strength of the greedy baseline.

\section{Experiments}

\label{sec:datasets_setup}
A practical challenge in this work is that the public data needed for sequential gym recommendation is largely missing. Ideally, such data would combine an exercise catalog, session-level order, prescription variables, user context, and interaction signals such as skipping or adherence. In practice, public gym-related data is fragmented: exercise catalogs describe movements, workout logs and plan templates often lack interaction context, and sensor datasets usually target recognition or form assessment rather than recommendation.

For that reason, this work combines structured exercise content, StrengthLevel-derived load references, synthetic users, and a simulator. Synthetic users include age, sex/gender, height, weight/BMI, goal, experience, training frequency, and capacity. The simulator is a controlled proxy rather than a full model of gym behavior: it is grounded in variables that matter for prescription, including user capacity, accumulated workload, muscle involvement, routine fit, and session history. This makes it suitable for comparing methods under data scarcity, but not for claiming clinical validity.

Skip-enabled environments model a weak but realistic interaction signal: users often do not rate every recommendation, but they may accept, skip, or abandon it. At each step, the simulator samples a binary skip outcome from a logistic model driven mainly by accumulated workload relative to user capacity, with additional effects from muscle strain, routine mismatch, recency, step position, and a base skip rate. Skip-only feedback updates online estimates of skip bias, capacity, and per-muscle avoidance, which are then included in later observations.

All environments operate on a filtered catalog with $N=217$ exercises. For full-prescription environments, sets, repetitions, and load are discretized into fixed bins: sets $(2,3,4,5,6)$, repetitions $(2,3,\dots,20)$, and 21 load bins over $[0.20,1.20]$ of a baseline one-repetition maximum (1RM). Load bins are converted to working weight using StrengthLevel lookups and the inverted Epley relation~\cite{Epley1985PoundageChart,LeSuer1997AccuracyPredictionEquations, strengthlog_exercise_directory}.

All RL agents are trained using Proximal Policy Optimization (PPO)  \cite{schulman2017ppo} with the same architecture and hyperparameters across runs: a Multi-Layer Perceptron (MLP) policy with hidden sizes $[256,256]$, learning rate $3\cdot10^{-4}$, rollout length 2048, batch size 256, discount factor $\gamma=0.99$, entropy coefficient 0.01, and clip range 0.2. We evaluate policies by rolling out complete episodes and recording return, reward components, and, in skip environments, completion and skipped ratio. Main results use continual evaluation so that user history and updates accumulate naturally.

\paragraph{Baselines}
To test whether RL is necessary, PPO is compared against three baselines. The random baseline samples uniformly from the action space. The greedy baseline constructs a session sequentially using a hand-designed score based on routine alignment, uniqueness, local diversity, goal hints, persistence, strain avoidance, and tie-breaking noise. This makes greedy a strong and partly reward-aligned baseline: its exercise-only performance should therefore be interpreted as evidence that a hand-designed policy can already solve much of the simpler sequencing task.

In full-prescription environments, greedy uses fixed prescription rules for sets, repetitions, and load based mainly on goal and experience. This makes it less adaptive to the coupled exercise-and-dose problem, where a plausible session requires both good exercise sequencing and user-appropriate prescription. The Particle Swarm Optimization (PSO) baseline performs offline planning for each episode using an integer variant of PSO, evaluating candidate plans with the terminal reward. Unlike PPO, PSO commits to a complete plan and cannot adapt online to skip feedback during an episode.

\paragraph{Evaluation metrics}
The main evaluation metric is mean return. For the skip-enabled environments, we additionally inspect completion ratio and skipped ratio. To avoid over-interpreting small differences, mean returns are reported together with 95\% bootstrap confidence intervals in the evaluation plots. Full confidence intervals, component-level results, and the corresponding bootstrap confidence-interval comparisons are provided in the supplementary material.

\paragraph{Reproducibility}
Our implementation, code, and supplementary materials are available at \url{https://anonymous.4open.science/r/RL-RecSys-for-Gym-Exercises-FINAL-anon/}.

\section{Results}

\begin{figure}[t]
    \centering
    \includegraphics[width=\columnwidth]{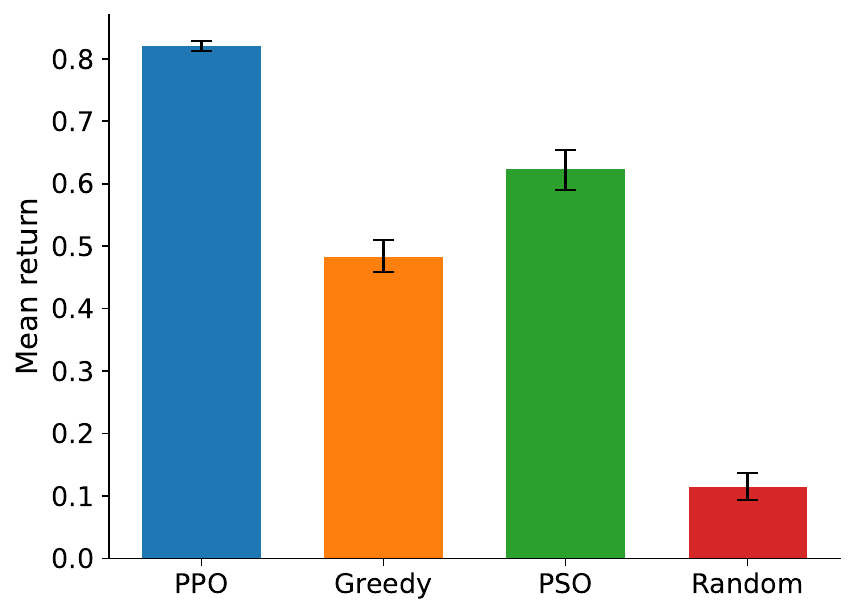}
    \vspace{-10pt}
    \caption{Baseline comparison in the full-prescription skip setting using mean return $\pm$ 95\% bootstrap confidence intervals.}
    \label{fig:fps_return_ci}
\end{figure}

Figure~\ref{fig:fps_return_ci} shows the most complete setting, FPS, where the recommender must jointly choose exercises, sets, repetitions, and load while adapting to skip feedback. PPO achieves the highest mean return and is separated from the baselines, supporting the main claim that RL becomes most useful once prescription and interaction are modeled together,

Table~\ref{tab:main_results} extends this comparison across all environments and user-pool regimes. In the exercise-only environments, greedy performs best, random performs worst, and PPO remains competitive with PSO. This should not be interpreted as a failure of RL. Rather, the exercise-only task is simple enough for a reward-aligned hand-designed policy: greedy directly encodes routine alignment, diversity, uniqueness, and strain avoidance, which are also important reward components.

The ranking changes in the full-prescription environments, where PPO outperforms all baselines. In the static pool, PPO improves over greedy by $0.389$ return in FPNS and by $0.338$ return in FPS, while also outperforming PSO by $0.206$ and $0.198$, respectively. The mechanism is that full prescription couples exercise selection with dose selection. A session can have good exercise structure but still receive a low score if sets, repetitions, or load are implausible for the user. Greedy remains effective at local sequencing, but its fixed prescription rules do not adapt enough to hidden capacity and accumulated workload. PSO searches complete plans, but in skip-enabled settings it cannot adapt online during the episode.

The skip-enabled environments show that interaction changes the optimization problem itself: no-skip settings reward sessions that look good on paper, whereas skip settings reward sessions that are actually completed. Skip-enabled models therefore become more adherence-aware, but optimizing completion can trade off against other components, such as fatigue or prescription quality. Finally, PPO reaches similar final performance under static and dynamic user pools, while chaotic, rapidly reversing drift makes skip-based personalization less reliable.

\begin{table}[t]
\centering
\caption{Mean return of PPO and the baselines in the realistic static and dynamic pool regimes. Higher values indicate better overall performance. Bold indicates the best mean in each row. $^\dagger$ indicates that the best method is separated from the strongest alternative in the corresponding 95\% bootstrap confidence-interval analysis reported in the supplementary material.}
\label{tab:main_results}
\begin{tabular}{llcccc}
\toprule
Environment & Pool & PPO & Greedy & PSO & Random \\
\midrule
EONS & Static & 0.859 & \textbf{0.932}$^\dagger$ & 0.849 & 0.563 \\
EONS & Dynamic & 0.863 & \textbf{0.932}$^\dagger$ & 0.849 & 0.563 \\
EOS & Static & 0.794 & \textbf{0.883}$^\dagger$ & 0.799 & 0.475 \\
EOS & Dynamic & 0.805 & \textbf{0.883}$^\dagger$ & 0.799 & 0.475 \\
FPNS & Static & \textbf{0.889}$^\dagger$ & 0.500 & 0.683 & 0.077 \\
FPNS & Dynamic & \textbf{0.892}$^\dagger$ & 0.500 & 0.683 & 0.077 \\
FPS & Static & \textbf{0.821}$^\dagger$ & 0.483 & 0.623 & 0.115 \\
FPS & Dynamic & \textbf{0.823}$^\dagger$ & 0.483 & 0.623 & 0.115 \\
\bottomrule
\end{tabular}
\end{table}

\section{Conclusion}
This paper examined whether reinforcement learning improves personalized workout recommendation when the problem is extended from exercise selection to realistic, gym-based full prescription. Building on the HFRL framework of~\citet{keeping_ijcai2023p692}, we adapted a home-fitness exercise recommender to a gym setting, incorporating exercise selection, sets, repetitions, load, skip behavior, online personalization, and evolving user characteristics.

Our results show that the benefit of reinforcement learning depends critically on the structure of the recommendation problem. In exercise-only environments, PPO consistently outperformed weaker baselines but did not surpass the strong greedy baseline — an important boundary result suggesting that, when the task is limited to exercise sequencing, a well-designed hand-crafted policy can capture much of the reward structure. In full-prescription environments, however, PPO emerged as the strongest method, outperforming all baselines including greedy. This is the paper's central finding: the advantage of RL manifests when the recommender must jointly solve exercise sequencing and dosage under user-specific capacity constraints.

The skip-enabled environments further reveal that realistic interaction is not merely an evaluation detail. When users may skip recommendations, the system optimizes for completed sessions rather than idealized ones, producing policies that are more adherence-aware. This comes with trade-offs, however, as optimizing for completion can tension with other reward components. Together, these findings suggest that RL is most valuable for the sequential, personalized, and interaction-dependent aspects of fitness recommendation.

An important avenue for future work is validating the framework with real users, richer feedback signals, and longer-term training plans. A user study evaluating the proposed framework under realistic, full-prescription conditions would be particularly valuable, as it would directly expose the gap between simulation and real-world gym behavior including factors such as equipment availability, pain, motivation, time pressure, and social context. Extending the reward function and prescription model beyond hand-designed proxy components, for instance through data-driven calibration or user feedback, would further strengthen the framework's applicability in practice.

\section*{GenAI Usage Disclosure}
The authors used OpenAI's ChatGPT to assist with shortening and restructuring thesis text into a short-paper draft, improving grammar and clarity, and suggesting fixing spelling errors. The tool was not used to generate experimental results, run experiments, fabricate data, or make autonomous scientific decisions. All claims, citations, tables, figures, and final text were reviewed and verified by the authors, who take full responsibility for the content of the paper.

\bibliographystyle{ACM-Reference-Format}
\balance
\bibliography{ref}

\end{document}